\documentclass[11pt]{article}
\usepackage{amsfonts,amssymb, bm, graphicx,a4,mdframed}

\newtheorem{lema}{Lemma}[section]
\newtheorem{teo}{Theorem}[section]
\newtheorem{defi}{Definition}[section]
\newtheorem{pro}{Proposition}[section]

\def\R{\mathbb{R}}
\def\Rd{{\mathbb{R}}^d}
\def\0{\emptyset}

\begin{document}


\voffset=-1.5truecm\hsize=16.5truecm    \vsize=24.truecm
\baselineskip=14pt plus0.1pt minus0.1pt \parindent=12pt
\lineskip=4pt\lineskiplimit=0.1pt      \parskip=0.1pt plus1pt

\def\ds{\displaystyle}\def\st{\scriptstyle}\def\sst{\scriptscriptstyle}


\let\a=\alpha \let\b=\beta \let\chi=\chi \let\d=\delta \let\e=\varepsilon
\let\f=\varphi \let\g=\gamma \let\h=\eta    \let\k=\kappa \let\l=\lambda
\let\m=\mu \let\n=\nu \let\o=\omega    \let\p=\pi \let\ph=\varphi
\let\r=\rho \let\s=\sigma \let\t=\tau \let\th=\vartheta
\let\y=\upsilon \let\x=\xi \let\z=\zeta
\let\D=\Delta \let\F=\Phi \let\G=\Gamma \let\L=\Lambda \let\Th=\Theta
\let\O=\Omega \let\P=\Pi \let\Ps=\Psi \let\Si=\Sigma \let\X=\Xi
\let\Y=\Upsilon

\global\newcount\numsec\global\newcount\numfor
\gdef\profonditastruttura{\dp\strutbox}
\def\senondefinito#1{\expandafter\ifx\csname#1\endcsname\relax}
\def\SIA #1,#2,#3 {\senondefinito{#1#2}
\expandafter\xdef\csname #1#2\endcsname{#3} \else \write16{???? il
simbolo #2 e' gia' stato definito !!!!} \fi}
\def\etichetta(#1){(\veraformula)
\SIA e,#1,(\veraformula)
 \global\advance\numfor by 1
 \write16{ EQ \equ(#1) ha simbolo #1 }}
\def\etichettaa(#1){(A\veraformula)
 \SIA e,#1,(A\veraformula)
 \global\advance\numfor by 1\write16{ EQ \equ(#1) ha simbolo #1 }}
\def\BOZZA{\def\alato(##1){
 {\vtop to \profonditastruttura{\baselineskip
 \profonditastruttura\vss
 \rlap{\kern-\hsize\kern-1.2truecm{$\scriptstyle##1$}}}}}}
\def\alato(#1){}
\def\veroparagrafo{\number\numsec}\def\veraformula{\number\numfor}
\def\Eq(#1){\eqno{\etichetta(#1)\alato(#1)}}
\def\eq(#1){\etichetta(#1)\alato(#1)}
\def\Eqa(#1){\eqno{\etichettaa(#1)\alato(#1)}}
\def\eqa(#1){\etichettaa(#1)\alato(#1)}
\def\equ(#1){\senondefinito{e#1}$\clubsuit$#1\else\csname e#1\endcsname\fi}
\let\EQ=\Eq


\def\bb{\hbox{\vrule height0.4pt width0.4pt depth0.pt}}\newdimen\u
\def\pp #1 #2 {\rlap{\kern#1\u\raise#2\u\bb}}
\def\hhh{\rlap{\hbox{{\vrule height1.cm width0.pt depth1.cm}}}}
\def\ins #1 #2 #3 {\rlap{\kern#1\u\raise#2\u\hbox{$#3$}}}
\def\alt#1#2{\rlap{\hbox{{\vrule height#1truecm width0.pt depth#2truecm}}}}

\def\pallina{{\kern-0.4mm\raise-0.02cm\hbox{$\scriptscriptstyle\bullet$}}}
\def\palla{{\kern-0.6mm\raise-0.04cm\hbox{$\scriptstyle\bullet$}}}
\def\pallona{{\kern-0.7mm\raise-0.06cm\hbox{$\displaystyle\bullet$}}}


\def\data{\number\day/\ifcase\month\or gennaio \or febbraio \or marzo \or
aprile \or maggio \or giugno \or luglio \or agosto \or settembre \or
ottobre \or novembre \or dicembre \fi/\number\year}

\setbox200\hbox{$\scriptscriptstyle \data $}

\newcount\pgn \pgn=1
\def\foglio{\number\numsec:\number\pgn
\global\advance\pgn by 1}
\def\foglioa{a\number\numsec:\number\pgn
\global\advance\pgn by 1}



\def\sqr#1#2{{\vcenter{\vbox{\hrule height.#2pt
\hbox{\vrule width.#2pt height#1pt \kern#1pt \vrule
width.#2pt}\hrule height.#2pt}}}}
\def\square{\mathchoice\sqr34\sqr34\sqr{2.1}3\sqr{1.5}3}

\let\ciao=\bye \def\fiat{{}}
\def\pagina{{\vfill\eject}} \def\\{\noindent}
\def\bra#1{{\langle#1|}} \def\ket#1{{|#1\rangle}}
\def\media#1{{\langle#1\rangle}} \def\ie{\hbox{\it i.e.\ }}
\let\ii=\int \let\ig=\int \let\io=\infty

\let\dpr=\partial \def\V#1{\vec#1} \def\Dp{\V\dpr}
\def\oo{{\V\o}} \def\OO{{\V\O}} \def\uu{{\V\y}} \def\xxi{{\V \xi}}
\def\xx{{\V x}} \def\yy{{\V y}} \def\kk{{\V k}} \def\zz{{\V z}}
\def\rr{{\V r}} \def\zz{{\V z}} \def\ww{{\V w}}
\def\Fi{{\V \phi}}

\let\Rar=\Rightarrow
\let\rar=\rightarrow
\let\LRar=\Longrightarrow

\def\lh{\hat\l} \def\vh{\hat v}

\def\ul#1{\underline#1}
\def\ol#1{\overline#1}

\def\ps#1#2{\psi^{#1}_{#2}} \def\pst#1#2{\tilde\psi^{#1}_{#2}}
\def\pb{\bar\psi} \def\pt{\tilde\psi}

\def\E#1{{\cal E}_{(#1)}} \def\ET#1{{\cal E}^T_{(#1)}}
\def\LL{{\cal L}}\def\RR{{\cal R}}\def\SS{{\cal S}} \def\NN{{\cal N}}
\def\HH{{\cal H}}\def\GG{{\cal G}}\def\PP{{\cal P}} \def\AA{{\cal A}}
\def\BB{{\cal B}}\def\FF{{\cal F}}

\def\tende#1{\vtop{\ialign{##\crcr\rightarrowfill\crcr
              \noalign{\kern-1pt\nointerlineskip}
              \hskip3.pt${\scriptstyle #1}$\hskip3.pt\crcr}}}
\def\otto{{\kern-1.truept\leftarrow\kern-5.truept\to\kern-1.truept}}
\def\arm{{}}
\font\bigfnt=cmbx10 scaled\magstep1
\def\ind#1{\1_{\{#1\}}}
\def\TT{{\mathcal{T}}}
\def\1{\rlap{\mbox{\small\rm 1}}\kern.15em 1}
\newcommand{\card}[1]{\left|#1\right|}

\BOZZA
\def\La{\Lambda}
\title{Classical particles in the continuum  subjected\\ to high density  boundary conditions}
\author{\normalsize
Aldo Procacci\thanks{\scriptsize Departamento de Matem{\'a}tica,
Universidade Federal de Minas Gerais, Belo Horizonte-MG, Brazil -
aldo@mat.ufmg.br}
~and
Sergio A. Yuhjtman\thanks{\scriptsize Departamento de Matem\'atica,
Universidad de Buenos Aires, Buenos Aires, Argentina -
syuhjtma@dm.uba.ar}}

\maketitle

\begin{abstract}
We  consider a continuous system of  classical particles
confined in a finite region $\L$ of $\mathbb{R}^d$
interacting through a superstable and tempered pair potential in
presence of non free boundary conditions. We prove that  the thermodynamic limit  of the pressure of the system
at any fixed inverse temperature  $\b$ and any fixed fugacity $\l$ does not depend on boundary conditions produced by  particles outside $\L$
whose density may increase sub-linearly with the distance  from the origin at a  rate which depends on how fast the pair potential decays at
large distances.  In particular,    if  the pair potential $v(x-y)$ is of Lennard-Jones type, i.e. it  decays as $C/\|x-y\|^{d+p}$ (with $p>0$) where
$\|x-y\|$ is the Euclidean distance between $x$ and $y$, then the existence of the thermodynamic limit of the pressure is guaranteed in presence of boundary conditions generated by external particles which may be distributed with a density increasing with the distance $r$ from the origin as $\r(1+ r^q)$, where $\r$ is any positive constant  (even arbitrarily larger than the density $\r_0(\b,\l)$ of the system evaluated with free boundary conditions) and  $q\le {1\over 2}\min\{1, p\}$.

\end{abstract}

\section{Introduction}
In the area of rigorous results in  statistical mechanics
it is a widely accepted  belief  that the entropy, free energy and pressure of a many-body system
are  independent on the boundary conditions imposed to the system. This is a well established fact for
bounded spin systems in a lattice
and simple proofs  can be found in many elementary textbooks. The situation becomes less clear
when unbounded spin systems on a lattice are analyzed. In that case the proofs of  the independency of the free energy
from the boundary conditions are much more involved and in general limitations  on the allowed
boundary conditions are needed, see e.g. \cite{LP}, \cite{COP} and \cite{PS}. The situation is  even less clear when
we consider continuous systems constituted  by many classical particles confined in a box $\L$ of the $d$-dimensional Euclidean space $\mathbb{R}^d$  and
interacting via a  pair potential (e.g.
the Lennard-Jones potential or other similar potentials).
Concerning specifically these systems, the vast majority of the rigorous result about the properties of the thermodynamic functions (e.g. pressure, free energy, entropy) in the thermodynamic limit
 have been deduced (mainly in the sixties/seventies, but also  more recently) in ensembles submitted to free boundary  condition or periodic
boundary conditions. We refer the reader in particular to the  papers  \cite{Pe63}, \cite{Ru63}, \cite{Ru63b}, \cite{LPe}, \cite{Do}, \cite{F}, \cite{DM}, \cite{Ru70}, \cite{FL},  to the overlooked but relevant papers \cite{Ba1} and \cite{Ba2} and their recent revisitation \cite{dLPY}, to the classic books \cite{Ru},  \cite{Ga}, and to \cite{PY}, \cite{NF} for  recent significative improvements. \newpage
\\There have been also some results  about  statistical ensembles of classical continuous particles subjected to boundary conditions
generated by external particles, mainly in regard to  the existence (and possible uniqueness) of the infinite volume Gibbs measure
 (see e.g.  \cite{PZ},\cite{KKP}, \cite{KPR}).

\\The only rigorous treatment we are aware of  concerning the thermodynamic limit of the pressure in a system of classical particles in the continuum subjected to non trivial boundary conditions has been given by Georgii in  \cite{Ge} (see also \cite{Ge2})
 where the independence of the pressure from a specific class of external  boundary conditions, called there  ``tempered" boundary conditions (see below),
 is proved under the assumption that the pair potential is superstable, regular  and diverging in a non summable way at the origin.

  \\In the present paper  we  somehow extend the results obtained by Georgii in \cite{Ge, Ge2}
   by proving  the existence of the thermodynamic limit of the pressure
   under a class of boundary conditions whose density is allowed to increase arbitrarily as one moves away from the origin.
To do this we make basically the same assumptions  on the  pair potential made by Georgii but we do not need to require
 that the potential must diverge in a non summable way at the origin.

\\More specifically,  we show that under the sole hypothesis
that the potential is superstable and regular, the thermodynamic limit of the pressure of a system of classical particles in the grand canonical ensemble at any fixed inverse temperature
$\b$ and any fixed activity $\l$ is independent of boundary conditions  produced by  particles outside $\L$
whose density may increase sub-linearly with the distance from the origin at a  rate which depends on how fast the pair potential decays at
large distances.   In particular,    if  the pair potential $v(x-y)$ is of Lennard-Jones type, i.e. it  decays as $C/\|x-y\|^{d+p}$ (with $p>0$) where
$\|x-y\|$ is the Euclidean distance between $x$ and $y$, then the existence of the thermodynamic limit of the pressure is guaranteed in presence of boundary conditions generated by external particles which may be distributed with a density increasing with the distance $r$ from the origin as $\r(1+ r^q)$, where $\r$ is any positive constant  (even hugely larger than the density $\r_0(\b,\l)$ of the system evaluated with free boundary conditions) and  $q\le {1\over 2}\min\{1, q\}$.

\section{Model and results}\label{sec2}
\numsec=2\numfor=1
\subsection{Model}

We consider a continuous system of classical particles confined in a
bounded compact region $\L$ of $\mathbb{R}^d$, which we assume  to be a cubic box of size $2L$ centered at the origin. So from now on
the symbol
$\lim_{\L\uparrow\infty}$  means simply that $L\to \infty$.
We denote by $x_i\in \mathbb{R}^d$ the
position vector of the $i^{\rm th}$ particle of the system and by
$\|x_i\|$ its Euclidean norm.
We suppose that particles
interact via a
translational invariant pair potential $v: \Rd\to \R\cup\{+\infty\}$  and are subjected  to a
boundary condition $\o$ generated by particles in fixed positions outside $\L$.
The boundary condition $\o$ is a locally finite
set of points of $\mathbb{R}^d$ representing the positions
of fixed particles in $\mathbb{R}^d$.  Namely, $\o$
must be  a countable set of points in $\mathbb{R}^d$ (not necessarily distinct) such that for
any compact subset $C\subset \mathbb{R}^d$ it holds that $|\o\cap
C|<+\infty$ (here $|\o\cap C|$ denotes the cardinality of the set
$\o\cap C$). We call   $\O$ the space of all locally finite configurations of
particles in $\mathbb{R}^d$ and, given a cube  $\L\subset \Rd$, we denote by $\O_\L$  the set of all finite
configurations of
particles in $\L$.

\\As usual, we will suppose that
each particle inside $\L$, say at position $x\in \L$, feels the
effect of the boundary condition $\o$ through the field
generated by the particles of the configuration $\o$ which are in
$\L^c=\mathbb{R}^d\setminus\L$. Free boundary conditions correspond to the case $\o=\0$.
We are interested in  studying  the
behavior of the system in the limit $\L\uparrow\infty$
with  a given boundary condition $\o$ and how eventually this limit may be
influenced by $\o$, having in mind that, as the volume $\L$  invades
$\mathbb{R}^d$, the fixed particles of $\o$ entering in $\L$ are
disregarded and only those boundary particles outside $\L$
influence particles inside $\L$.
 We will  denote below by $|\L|=(2L)^d$ the volume of
$\L$ and by  $\partial \L$  the boundary of $\L$. We define,  for $x\in \L$,
$$
d_x^\L=\inf_{y\in
\partial\L}\|x-y\|
$$

\\In the suite we will frequently use the  following notation. Given a configuration $\o\in \O$,
a  function $f: \Rd\to \R\cup\{ +\infty\}$,
a cubic box $\L\subset \Rd$ and a point  $x\in \L$, we set
$$
E^f_\L(x, \o)=\sum_{y\in \o\cap \L^c}f(x-y)
$$

\\With this notation,
 for any fixed volume $\L$ and for any fixed
boundary condition $\o$,  the partition function of the system in the
grand canonical ensemble at  inverse temperature $\b\ge 0$ and
fugacity $\l\ge 0$ is given by
$$
\Xi^\o_{\La}(\b,\l)=\sum_{n=0}^{\infty}{\l^{n}\over n!} \int_\L
dx_1\dots \int_\L dx_n e^{-\b\left[\sum\limits_{1\le i< j\le
n}v(x_i-x_j)+\sum\limits _{i=1}^nE^{v}_\L(x_i,\o)\right]}\Eq(1.1)
$$
where in the series of the r.h.s. the $n=0$ term is equal to one and
$ E^{v}_\L(x,\o)=  \sum_{y \in \o\cap \L^c} v(x-y)$
represents the field  felt by a particle
sitting in the point $x\in \L$ due to the fixed particles of the
boundary condition $\o$ located at points outside $\L$.

\\The finite volume pressure of the system is then given by
$$
\b p_\L^\o(\b,\l)={1\over |\L|}\log \Xi^\o_{\La}(\b,\l)\Eq(pressureb)
$$
and the thermodynamic limit of the finite volume pressure (if it exists)  is
$$
\b p^\o(\b,\l)=\lim_{\L\uparrow\infty}{1\over |\L|}\log \Xi^\o_{\La}(\b,\l)\Eq(prlim)
$$
\\The r.h.s. of \equ(1.1) is for the time being just a formal series and consequently r.h.s. of \equ(pressureb) and \equ(prlim)
are, for the time being, meaningless. The  well definiteness and the convergence of the series  in the  r.h.s. of \equ(1.1) depends on assumptions on the pair potential $v$ and on the boundary condition $\o$.

\\If we suppose  that $\o=\0$ (i.e. if we use  free boundary conditions), the term $\sum_{i=1}^nE^{v}_\L(x_i,\o)$ in the exponential
of the integrand of the r.h.s. of \equ(1.1) vanishes and
thus, the series
$$
\Xi^\0_{\La}(\b,\l) =\sum_{n=0}^{\infty}{\l^{n}\over n!} \int_\L
dx_1\dots \int_\L dx_n e^{-\b\sum\limits_{1\le i< j\le
n}v(x_i-x_j)}\Eq(1.1free)
$$
representing the partition function of the system subjected to free boundary conditions is such that
for any $n\in \mathbb N$   its $n^{th}$  coefficient, i.e. the integral
$$
\int_\L
dx_1\dots \int_\L dx_n e^{-\b\sum\limits_{1\le i< j\le
n}v(x_i-x_j)} \Eq(coeff)
$$
is  well defined (i.e. is  finite) just by imposing that $v$ takes values in $\R\cup \{{+\infty}\}$. It is long known (see e.g. \cite{Ru}) that
 the series \equ(1.1free) is  convergent if  the pair potential is stable according to the following definition.
\begin{defi}
A pair potential $v$ is stable if
there exists $B\ge 0$ such that
for all $n\in \mathbb{N}$ and for all $(x_1,\dots,x_n)\in
\mathbb{R}^{dn}$
$$
\sum_{1\le i<j\le n} v(x_i-x_j)\ge -B n\Eq(2.6)
$$
and the smallest constant $B$ satisfying \equ(2.6) is called the {\it stability constant} of the potential.
\end{defi}
In particular it is immediate to see that, if \equ(2.6) holds,  the $n$-order coefficient  \equ(coeff) is bounded
by $|\L|^ne^{n\b B}$ and therefore the series \equ(1.1free) is  an analytic function of $\l$ for all $\l\in \mathbb C$.

\\The existence of the infinite-volume pressure even when  free boundary conditions are adopted
$$
p^\0(\b,\l)=\lim_{\L\to \infty}p_\L^\0(\b,\l)= \b^{-1} \lim_{\L\to \infty}{1\over |\L|}\log \Xi^\o_{\La}(\b,\l)\Eq(poo)
$$
is a non trivial issue   and to prove it one has to do some further assumptions on the pair potential.  Let us give the following definitions.
\begin{defi}
\\A pair potential $v(x)$ is regular if
$$
\int_{\Rd} |e^{-\b v(x)}-1|dx<+\infty \Eq(regu)
$$
\end{defi}
As   shown in \cite{Ru63} and \cite{Pe63} the finiteness of the integral given in \equ(regu) guarantees
that the zero-free region  of the partition function \equ(1.1free) around $\l=0$  in the complex plane
 does not shrink to zero as $\L\to \infty$. This implies the existence of
a disc $D_R=\{\l\in \mathbb C: |\l|\le R\}$ with $R$ independent of $\L$ in which
$\log\Xi^\0_{\La}(\b,\l)$ is analytic (see e.g. Chapter 4 in \cite{Ru}).  If the model has to describe a real gas this
is a minimal request: at least for small values of the fugacity the system must be a pure gas.

\begin{defi}
A pair potential $v$ is {\it superstable} if $v$ can be written as
$$
v=v_1+v_2
$$
with $v_1$ {\it  stable}
and  $v_2$ non-negative and strictly positive near the origin.
\end{defi}
The existence of the limit \equ(poo) and its continuity as a function of $\l$ and $\b$, when particles interact via a  superstable and regular pair potential is a well established fact since the sixties (see \cite{Do}, \cite{F}, \cite{Ru63b}, \cite{DM}, \cite{Ru70} and \cite{Ru}).  Much later   Georgii \cite{Ge,Ge2} showed
  that the limit \equ(prlim) with non free boundary conditions exists if the pair potential, beyond superstable and regular,  has a hard-core or diverges in a non summable way at short distances. Its result holds for all ``tempered boundary conditions $\o$ (see (2.24) in \cite{Ge}  or (2.6) in \cite{Ge2}), which basically means that, for some finite positive constant $t$, $\o$ must be such that $\limsup_{\La\to \infty} |\L|^{-1}\sum_{\D_\d\in \L_\d}|\o\cap \D_\d|^2\le t$ where $\L_\d$ is a collection of cubes $\D_\d$ of fixed size $\d>0$ forming a partition of $\L$.

  \\In the present paper the  assumptions on the pair potential and on the allowed boundary conditions are as follows.
\subsubsection{Assumptions on the pair potential}
\vskip.1cm
\\The translational invariant pair potential $v(x)$ is supposed to be   Lebesgue measurable and to satisfy the
following assumptions.
\vskip.2cm
\begin{itemize}
\item[(i)]   $v$ is {\it superstable}. Namely, $v$ can be written as the sum of two functions
$$
v=v_1+v_2
$$
with $v_1$ {\it  stable} with stability constant $B$
and  $v_2$ non-negative and strictly positive near the origin in a strong sense: there exist
two constants  $a>0$ and $c>0$ such that
$$
v_2(x)\ge c~~~~~~~~~~~~~~~\mbox{for all $\|x\|\le a$}  \Eq(deca)
$$
\\
\item[(ii)]   $v$ is {\it tempered}, namely,
there exist $b>0$ and a  non-negative monotonic  decreasing function $\h: [0,+\infty)\to [0,+\infty)$ \index{pair
potential!temperness} such that,
$$
\int_{0}^\infty\h(r)r^{d-1}dr< \infty \Eq(2.7b)
$$
and
$$
|v(x)|\le \h(\|x\|)~~~~~~~~~~~~~~~~~~\mbox{for all $x\in \Rd$ such that $\|x\|\ge b$}\Eq(2.7)
$$
\end{itemize}
\vskip.2cm
\\Note that assumption (ii) is basically equivalent to impose that $v$ is regular according \equ(regu) (see comment after Definition 4.1.2 in \cite{Ru}).
\\Let us  define for later convenience
$$
v^\pm(x)= \max\{0, \pm v(x)\}
$$
so that
$$
v(x)= v^+(x)-v^-(x)
$$
\\Assumption (ii) immediately implies that $v^-$ and $v^+$ are  such that $v^\pm(x)\le \eta(\|x\|)$  for all
$x\in \Rd$ such that $\|x\|\ge b$. Moreover, due to stability  $v^-$ is bounded {\it tout court}
in the whole space  $\Rd$, i.e., $v^-(x)\le 2B$ for all $x\in \Rd$.

\vskip.2cm
\\{\bf Remark}
\\We will  assume, without loss of generality, that $\eta$ is a  continuous function which is constant in the interval $[0, b]$ at the  value $2B$ (i.e.  $\h(r)= 2B$ for all $r\le b$) so that
$v^-(x)\le \h(\|x\|)$ for all $x\in \Rd$. Moreover we will choose $\eta$   sufficiently well behaved in such a way that, for  $\d>0$  small enough and for any
cube $\D\subset \mathbb{R}^d$ of size $\d$, there is a constant $C_\d$ independent on the position of $\D$ in $\mathbb{R}^d$ such that
$$
\d^{d}\sup_{y\in \D_\d}\eta(\|y\|)\le C_\d \int_{y\in \D_\d} \eta(\|y\|)dy \Eq(kade)
$$
We will use inequality \equ(kade) in the following.

\vskip.2cm
\\We
 further define the function $V:  [0,+\infty)\to (0,+\infty)$ with
$$
V(r)=  \int_{\Rd\setminus B_r(0)}^\infty \eta(\|x\|)dx
\Eq(er)
$$
where  $B_r(0)=\{x\in \Rd: \|x\|< r\}$ is the $d$-dimensional open ball of radius $r$ centered at the origin in $\Rd$.
\\Note that, due to  \equ(2.7b), it holds that
$$
\lim_{r\to\infty} V(r)=0 \Eq(vrinfty)
$$

\\We now establish the class of boundary conditions under which the thermodynamic limit of the pressure can be proved to exist.
\def\Zd{{\mathbb{Z}}^d}
\subsubsection {The allowed boundary conditions}
\vskip.1cm

\\We will suppose hereafter
that $\R^d$ is partitioned in elementary cubes $\D_\d$ of suitable size $\d>0$. Along the paper we will  denote by $\R^d_\d$ the set of all these cubes and,  given $x\in \Rd$, we will denote by $\D_\d(x)$ the cube of $\Rd_\d$ to which $x$ belongs. Moreover,  given a
$d$-dimensional cube  $\L$  of size $2L$ centered at the origin of $\Rd$,
we agree to choose $\d$  in such a way that $2L/\d$ is integer  and we call $\L_\d$ the set whose elements are the elementary cubes  forming $\L$.

\\Given  $\o\in \O$, we  define the density of $\o$ as the function $\r_\d^\o: \Rd\to [0, +\infty): x\mapsto \r_\d^\o(x)$   with
$\r_\d^\o(x)=\d^{-d}|\o\cap \D_\d(x)|$ (hence $\r_\d^\o(x)$ is constant for all $x\in \D_d(x)$). Since $\o$ is locally finite, $\r_\d^\o(x)$ is everywhere finite.
 \begin{defi}\label{admi}
Given a superstable and tempered pair potential $v$ according to the assumptions {\rm (i)} and {\rm (ii)},
a continuous  monotonic non-decreasing  function  $g: [0,+\infty)\to [0,+\infty) $ is called admissible if the following conditions hold.

  $$
  g(\a+\b)\le g(\a)+g(\b)\Eq(subli)
  $$
$$
\int_{\Rd}\h(\|x\|)g(\|x\|)dx<+\infty \Eq(integr)
$$
An admissible function is called non-trivial if
$$
\lim_{u\to\infty}g(u)=+\infty\Eq(ntriv)
$$
\end{defi}

\\Let $\r$ be  a non-negative constant and let $g: [0,+\infty)\to [0,+\infty) $ be admissible.
We will set
$$
\O_{\r,g}=\{\o\in \O: \r_\d^\o(x)\le \r(1+ g(\|x\|)), \forall x\in \Rd\}\Eq(orog)
$$
and define   the set of allowed configurations as
$$
 \O^*_{g}=\cup_{\r\ge 0} \O_{\r,g}\Eq(oog)
$$
\\Therefore, the allowed configurations in  $\O^*_{g}$ are those
whose density increases at most as  $ \r(1+ g(\|x\|))$ for some constant $\r$, where $g$ is an admissible function
according to Definition
\ref{admi}. Note that if $g$ is non-trivial, the density $\r_\d^\o(x)$ of a configuration
$\o\in \O^*_g$ becomes arbitrarily large as we move away from the origin. On the other hand when $g$
is identically zero,  $\O^*_0$ is the set of configurations with bounded density.
\vskip.1cm
\\{\bf Remark}. It should be noted that although the intermediate set $\O_{\r,g}$ defined in \equ(orog) depends on $\d$, the set of allowed configurations $\O^*_g$ defined in \equ(oog) does not depend on the choice of $\d$.

\\Indeed,  consider two different partitions of $\mathbb{R}^d$ where
the cubes have sizes $\d_1$ and $\d_2$.
We take $\o \in \O_{\rho, g}^{\d_1}$ and we will show that, for some finite $\tilde \r$,  $\o \in
\O_{\tilde \rho, g}^{\d_2}$.
Let $x \in \Rd$ and consider all the cubes in the $\d_1$-partition that
intersect the cube of the $\d_2$-partition containing $x$. Take points
$y_1, ..., y_m$, one in each such cube.
Then, for some constant $K$ depending only on $\d_1$ and $\d_2$, we have:
$$\rho_{\d_2}^\o (x) \le \sum_{i=1}^m K \rho_{\d_1}^\o(y_i) \leq
\sum_{i=1}^m K \rho(1 + g(\|y_i\|))$$
For $\|x\|$ sufficiently large, say $\|x\|>R$, we have $\|y_i\| \le 2\|x\|$, so using
the properties of $g$ we have
$g(\|y_i\|) \le 2g(\|x\|)$. Thus, in these cases we have
$$\rho_{\d_2}^\o (x) \le mK\rho(1+2g(\|x\|)) \leq 2mK\rho (1 +
g(\|x\|)) = \rho' (1 + g(\|x\|)) $$
For the values of $x$ such that $\|x\|\le R$ we can simply pick
$\rho''>0$ such that
$$\rho_{\d_2}^\o (x) \le \rho'' \leq \rho''(1 + g(\|x\|))$$
Therefore if we  take $\tilde\r=\max\{\r', \r''\}$ we have that $\rho_{\d_2}^\o (x)\le \tilde \r(1 + g(\|x\|))$ for all $x \in \mathbb{R}^d$.
\vskip.5cm

\\We define, for later use, the function $W:[0,+\infty)\to [0,+\infty)$ such that for any $r\ge 0$,
$$
W(r)=\int_{\Rd\setminus B_r(0)} \eta(\|x\|)g(\|x\|)dx\Eq(wr)
$$
Note that, by \equ(integr), we have that
$$
\lim_{r\to \infty} W(r)=0\Eq(wrinfty)
$$

\\We will now show  that the assumptions on the pair potential and on the boundary conditions established above guarantee that
the grand canonical  partition function defined in \equ(1.1) is an analytic  function of $\l$ in the whole complex plane.

\\We first show the following preliminary Lemma.

\begin{lema}\label{p2}
Let $v$ be a pair potential satisfying assumptions {\rm(i)} and {\rm (ii)},  let $g$ be admissible and let $\o\in \O^*_g$.
Then there exists  a  finite constant $\tilde \k$  such that, for any $x\in\L$
$$
E^{v^-}_\L(x,\o)\le\tilde \k(1+ g({L}))
$$
\end{lema}
{\bf Proof}. If $\o\in \O^*_g$, then there exists $\r\in [0, \infty)$ such that  $\r_\d^\o(y)\le \r(1+ g(\|y\|))$ for all $y\in \Rd$. Then
given  $x\in \Rd$ and   $r\ge 0$  we have
$$
 E^{v^-}_{\L}(x,\o) = \sum_{y \in \o\atop y\in \L_c} v^-(x-y)\le \sum_{y \in \o} v^-(x-y)\le \sum_{\D_\d\in \Rd_\d} \sup_{y\in \D_\d}
 v^-(x-y)|\o\cap\D_\d|
 $$
 $$
 \le~
\d^d\sum_{\D_\d\in \Rd_\d}\sup_{y\in\D_\d}\eta(\|x-y\|)\r_\d^\o(y)~~~~~~~~~~~~~~~~~~~~~~~~~~~~~~~
$$
Now, by inequality \equ(kade), we have
$$
\d^{d}\sup_{y\in \D_\d}\eta(\|x-y\|)\r^\o_\d(y)\le C_\d \int_{y\in \D_\d} \eta(\| x- y\|)\r^\o_\d( y)dy
$$
Hence
\begin{eqnarray*}
E^{v^-}_{\L}(x,\o) &\le & C_\d\sum_{\D_\d\in \Rd_\d}
\int_{y\in \D_\d} \eta(\| x- y\|)\r^\o_\d( y)dy=C_\d\int_{\mathbb{R}^d} \eta(\| x- y\|)\r^\o_\d( y)dy\\
&= & C_\d\r\int_{\mathbb{R}^d}\eta(\|x-y\|)(1+g(\|y\|))dy\\
&\le & C_\d\r\int_{\mathbb{R}^d}\eta(\|x-y\|)(1+g(\|x-y\|+\|x\|))dy
\end{eqnarray*}
Therefore,
$$
E^{v^-}_{{\L}}(x,\o)\le C_\d \r  \left[\int_{\Rd}{g(\|y\|)\h(\|y\|)}dy+ (1+g(\|x\|))
\int_{\Rd}{\h(\|y\|)}dy\right]
$$
Recalling definitions \equ(er) and \equ(wr),
and observing that, for any  $x\in \L$ we have that $\|x\|\le \sqrt{d}L$ and that, by \equ(subli), $g(\sqrt{d}L)\le  g({d}L)\le dg(L)$,
we can conclude that
$$
E^{v^-}_{{\L}}(x,\o)\le   C_\d \r \left[W(0)+ (1+  dg(L))V(0)\right]\le \tilde \k(1+ g({L}))
$$
with  $\tilde \k =C_\d \r d(W(0)+V(0))$. $\Box$
\vskip.2cm
\\Lemma \ref{p2} above implies straightforwardly the following Proposition.
\begin{pro}\label{anpa}
Let $v$ be a pair potential satisfying assumptions {\rm(i)} and {\rm (ii)},  let $g$ be admissible and let $\o\in \O^*_g$.
Then the grand canonical  partition function defined in \equ(1.1) is an analytic  function of $\l$ in the whole complex plane.
\end{pro}

\\{\bf Proof}. By assumption (i) on the pair potential $v$ we have
$$
\sum\limits_{1\le i< j\le n}v(x_i-x_j)\ge  \sum\limits_{1\le i< j\le n}v_1(x_i-x_j) \ge -nB
$$
and, by Lemma \ref{p2},  we have  that
$$
E^v_\L(x,\o)\ge -\tilde \k (1+g(L)).
$$
Therefore, for any $\l\in \mathbb{C}$
$$
\Xi^\o_{\La}(\b,\l)\le \exp\Big\{|\l||\L| e^{\b B}e^{\b \tilde \k (1+g(L)) }\Big\}
$$
$\Box$

\\As a consequence of Proposition \ref{anpa}, considering that $\Xi^\o_{\La}(\b,\l)\ge 1$ when $\b\ge 0$ and $\l\ge 0$, we also have that the finite volume pressure \equ(pressureb) is
well defined and finite for all $(\b,\l)\in [0, +\infty)\times [0, +\infty)$ as soon as the pair potential is stable and tempered according to
(i) and (ii) and $\o\in \O_g^*$.  Of course, even with $p^\o_\L(\l,\b)$ well defined for every finite $\L$ and for every $\o\in \O^*_g$, the problem of the existence
of the thermodynamic limit \equ(prlim) and its independency on $\o$ is another story.

\subsection{Results}
\\We conclude this section by enunciating the main results of this note in form of four Theorems. The first two theorems establish general conditions  under which, for any boundary condition
$\o\in \O^*_g$,  $\limsup_{\L\uparrow \infty}{|\L|^{-1}} \log \Xi^\o_{\La}(\b,\l)$ and $\liminf_{\L\uparrow \infty}{|\L|^{-1}} \log \Xi^\o_{\La}(\b,\l)$
 are bounded from above and from below by $\lim_{\L\uparrow \infty}{1\over |\L|} \log \Xi^\0_{\La}(\b,\l)$ respectively.

\begin{teo}\label{sergiosup}
Consider a continuous system of  classical particles
interacting through a superstable and tempered pair potential  $v$ according to assumptions {\rm(i)} and {\rm (ii)} and let
$\o\in \O^*_g$ with $g$ admissible according to Definition \ref{admi}.

\\Let $V$ and $W$ be the functions defined in \equ(er) and \equ(wr) respectively and  suppose that
$$
\lim_{R\to \infty} g(R) {\int_{0}^R  W(s)ds\over R}=\lim_{R\to \infty} [g(R)]^2{\int_{0}^R V(s)ds\over R}=0\Eq(diffi)
$$
Then, for any $\l\ge 0$ and  $\b\ge 0$ it holds
$$
\limsup_{\L\uparrow \infty}{1\over |\L|} \log \Xi^\o_{\La}(\b,\l)\le
 \lim_{\L\uparrow \infty}{1\over |\L|} \Xi^\0_{\La}(\b,\l) \Eq(limsup)
$$
\end{teo}

\\{\bf Remark}. Condition \equ(diffi)  basically imposes constraints to the possible growth of the function $g$ depending on how rapidly the potential decays at  large distances.

\begin{teo}\label{sergioinf}
Consider a continuous system of  classical particles
interacting through a superstable and tempered pair potential  $v$ according to assumptions {\rm(i)} and {\rm (ii)} and let
$\o\in \O^*_g$ with $g$ admissible according to Definition \ref{admi}.

\\Let $V$ be the function defined in \equ(er) and  suppose that there exists a continuous function
$h(L)$ such that    $\lim_{L\to \infty} h(L)=\infty$,  $\lim_{L\to \infty} h(L)/L=0$ and
$$
\lim_{L\to \infty} (1+g(L)) V(h(L))=0\Eq(exi)
$$
Then, for any $\l\ge 0$ and  $\b\ge 0$ it holds
$$
\liminf_{\L\uparrow\infty}{1\over |\L|} \log \Xi^\o_{\La}(\b,\l)\ge
 \lim_{\L\uparrow \infty}{1\over |\L|} \Xi^\0_{\La}(\b,\l) \Eq(liminf)
$$
\end{teo}
\\The proofs of Theorem \ref{sergiosup} and Theorem \ref{sergioinf} are given  in Sections 3 and 4 respectively.

\\The next two theorems, which follows   straightforwardly from Theorem \ref{sergiosup} and \ref{sergioinf}, produce two relevant examples in which the thermodynamic limit of the pressure under boundary conditions
belonging to $\O^*_g$ exists and it is equal to the free boundary condition pressure.
\begin{teo}
Let $v$ superstable and tempered according to assumptions {\rm (i)} and {\rm (ii)}  and let $\o\in\O^*_0$  (i.e configurations with bounded density),
then
$$
\lim_{\L\uparrow\infty} \log \Xi^\o_{\La}(\b,\l)=
 \lim_{\L\uparrow \infty}{1\over |\L|} \Xi^\0_{\La}(\b,\l) \Eq(limlim)
$$
\end{teo}
\\{\bf Proof}. Let us first show that if $g=0$ (hence we are considering boundary conditions with bounded density),   any superstable and tempered pair potential satisfies \equ(diffi) and therefore, by Theorem \ref{sergiosup}, inequality \equ(limsup) holds. Indeed, if $g=0$ then the condition  \equ(diffi) simply boils down to
$$
\lim_{R\to \infty}{\int_{0}^RV(s)ds\over R}=0\Eq(faci)
$$
Now, if $\lim_{R\to \infty}\int_{0}^RV(s)ds<+\infty$ then equation \equ(faci) is trivially true. On the other hand, if
$\lim_{R\to \infty}\int_{0}^RV(s)ds=\infty$, then by l'Hopital rule
$$
\lim_{R\to \infty}{\int_{0}^RV(R)ds\over R}= \lim_{R\to \infty} V(R)=0
$$
and thus we have that inequality \equ(limsup) holds.

\\Secondly, if $g=0$, we can choose $h(L)=\sqrt{L}$ which clearly satisfies
the hypothesis of Theorem \ref{sergioinf}. Therefore,
by Theorem \ref{sergioinf}, also inequality \equ(liminf) holds. $\Box$

\vskip.2cm

\\In the second  example, we suppose that $v$ is of Lennard-Jones type. In this case the density distribution
of the boundary condition is allowed to increase sublinearly with the distance from the origin.

\begin{teo}\label{coro2}
Let   $v$  superstable and tempered according to assumptions (i) and (ii) and suppose that the function $\h$
is such that, for some constant $C$ and some $p>0$,
$$
\h(r)\le {C\over r^{d+p}}   ~~~~~~~~~~~~~~~~~~~\mbox{for all $r\ge b$}  \Eq(poly)
$$
Let $\o\in \O^*_g$ with $g(r)=r^q$ and $q>0$ such that
$$
q<{1\over 2}\min\{1, p\} \Eq(qsub)
$$
Then \equ(limlim) holds true.
\end{teo}
\\{\bf Proof}. Let us first prove that inequality \equ(limsup) holds by using Theorem \ref{sergiosup}. We start by showing that the
function $g(r)= r^q$  with $0<q<p$  is admissible according to Definition  \ref{admi}.
Clearly  $\lim_{r\to \infty} r^q=+\infty$ and, recalling that $\h(r)$ has been chosen to take the value $2B$ in the interval $[0,b]$,
$$
\int \h(\|x\|)g(\|x\|) dx \le 2B V_d b^{q+d}  + S_d\int_{b}^\infty {1\over r^{1+p-q}}dr <+\infty
$$
where  $V_d$  and $S_d$ are  the volume  and the surface of the $d$ dimensional unit sphere respectively.
Moreover  for any $q<1$ and any $\a,\b>0$ it holds
that $(\a+\b)^q\le \a^q+\b^q$.
In conclusion $g(r)=r^q$ is admissible for all $q$ such that $0<q<\min\{1, p\}$.

\\Now let us analyze the left hand side of \equ(diffi).
$$
\lim_{R\to\infty}  {g(R) {\int_{0}^R  W(s)ds}\over R}\Eq(lim1)
$$
In what follows we will denote as $K_1, K_2, \dots$ constants not depending on $R$ and  $S_d$ is the surface of the unit sphere in $d$ dimensions. Observe that
\begin{eqnarray*}
  \int_{0}^R  W(s)ds &=&\int_{0}^R ds \int_{\Rd\setminus B_s(0)} \|x\|^q\h(\|x\|)dx \\
   &=& \int_{0}^{b} ds \int_{\Rd\setminus B_s(0)} \|x\|^q\h(\|x\|)dx+ \int_{b}^R ds \int_{\Rd\setminus B_s(0)} \|x\|^q\h(\|x\|)dx \\
   &= & \int_{0}^{b} ds \int_{B_b(0)\setminus B_s(0)} \|x\|^q\h(\|x\|)dx+ \int_{0}^{b} ds \int_{\Rd\setminus B_b(0)} \|x\|^q\h(\|x\|)dx~+ \\
  &~& ~ + \int_{b}^R ds \int_{\Rd\setminus B_s(0)} \|x\|^q\h(\|x\|)dx \\
  &\le  & 2B \int_{0}^{b} ds \int_{B_b(0)\setminus B_s(0)} \|x\|^qdx+ \int_{0}^{b} ds \int_{\Rd\setminus B_b(0)}{C\over \|x\|^{d+p-q}}dx~+ \\
  &~& ~ + \int_{b}^R ds \int_{\Rd\setminus B_s(0)} {C\over \|x\|^{d+p-q}}dx \\
  \end{eqnarray*}
where in the last inequality here above we have used that, by assumption,   $\eta(\|x\|)=2B$ for $\|x\|< b$ and we have bounded
$\eta(\|x\|)\le{ C\over \|x\|^{d+p}}$  for $\|x\|\ge b$. Therefore we get

  \begin{eqnarray*}
\int_{0}^R  W(s)ds &\le& S_d\Bigg[2B\int_{0}^{b} ds \int_{s}^b r^{d-1+q} dr+ \int_{0}^{b}ds \int_{ b}^\infty {C\over r^{1+p-q}}dr
+ \int_{ b}^R ds \int_{s}^\infty {C\over r^{1+p-q}}dr\Bigg] \\
 &\le & S_d\Bigg[2B b \int_{0}^{ b} r^{d-1+q} dr+ \int_{0}^{ b}ds \int_{ b}^\infty {C\over r^{1+p-q}}dr~+~
 \int_{ b}^R ds \int_{s}^\infty {C\over r^{1+p-q}}dr\Bigg] \\
 &\le & S_d\Bigg[2B b\left( \int_{0}^{ b} r^{d-1+q} dr+ \int_{r}^\infty {C\over r^{1+p-q}}dr\right)
+ \int_{ b}^R ds \int_{s}^\infty {C\over r^{1+p-q}}dr\Bigg] \\
&\le& K_1+ S_d\int_{ b}^R {C\over s^{p-q}}ds
\end{eqnarray*}
Hence,
$$
\frac{g(R)}{R} \int_{0}^R  W(s)ds~\le ~K_1R^{q-1} +  CS_dR^{q-1} \int_{ b}^R{1\over s^{p-q}}ds
$$

\\Since $q < 1$, the number $R^{q-1}$ goes to zero when $R \to \infty$. Thus, to show \equ(diffi)
we only have to deal with $R^{q-1}\int_1^R{ s^{q-p}}ds$. Since $2q < p$, we can pick a $t > 0$ such that
$2q + t < p$, which is the same as $q-p < -q -t$. This $t$ can also be chosen so that $q + t < 1$.
$$
\int_1^R{s^{q-p}}ds < \int_1^R{s^{-q-t}}ds = {1 \over 1-q-t} (R^{1 - q - t} - 1 ) $$
$$R^{q-1} \int_1^R{s^{q-p}}ds < {1 \over 1-q-t}(R^{-t} - R^{q-1} )
\stackrel{R \to \infty}{\longrightarrow} 0 $$

\\The second term of \equ(diffi), namely
 $$
\lim_{R\to \infty} [g(R)]^2{\int_{0}^R V(s)ds\over R}\Eq(lim2)
$$
 can be analyzed proceeding similarly. Doing so we get
$$
{[g(R)]^2 \over R} \int_{0}^R V(s)ds\le K_2 R^{2q-1} + K_3 R^{2q-1}\int_{b}^R {1\over s^{p}}ds
$$
which goes to zero by imitating the above argument. This time we have to use $q < 1 / 2$ as well
as $q < p / 2$. This concludes the proof of inequality \equ(limsup).

\\Let us now  prove that also inequality \equ(liminf) holds. If $g(L)=L^q$ where $q<{1\over 2} \min\{1, p\}$ then we can choose e.g.
$h(L)=L^{2/3}$. Indeed with this choice we have clearly that $\lim_{L\to \infty}h(L)=+\infty$ and
 $\lim_{L\to \infty}{h(L)/L}=0$. Moreover recalling that by hypothesis
 $g(L)=L^q$, $V(L)\le C/L^p$ and $q <{1\over 2}\min\{1,p\}$, the l.h.s. of \equ(exi) is, for any $L>1$, such that
 $$
 (1+g(L)) V(h(L))\le {(1+CL^q)\over L^{{2\over 3}p}}={1\over L^{{2\over 3}p}}+ {C\over L^{{2\over 3}p-q}}
 $$
 and thus 
 $\lim_{L\to \infty} (1+g(L)) V(h(L))=0$.  Therefore, by Theorem \ref{sergioinf}
 inequality \equ(liminf) holds. In conclusion \equ(limlim) holds true if the hypothesis of Theorem \ref{coro2} stands. $\Box$

\vskip.2cm

\section{Proof of Theorem \ref{sergiosup}}
\\In   this section we will denote shortly by  $\vec x$   a generic configuration
$(x_1,...,x_n)\in \L^n$  so that  $\vec x\in \L$ means $(x_1,...,x_n)\in \L^n$ for some $n\in \mathbb{N}$. We will
use below the following shorter notations.
$$
\sum_{n=0}^\infty {\l^n\over n!} \int_\L
dx_1\dots \int_\L dx_n (\cdot)\doteq \int_{\O_\L}d\m_\l(\vec x)(\cdot)\Eq(poisson)
$$
$$
v(\vec x)= \sum_{\{x,y\}\subset \vec x} v(x-y)
$$
$$
E^{v}_\L(\vec x,\o)=\sum_{x\in \vec x}E^{v}_\L(x,\o)\Eq(evecx)
$$
$$
~E^{v^\pm}_\L(\vec x,\o)=\sum_{x\in \vec x}E^{v^\pm}_\L(x,\o)\Eq(evecpm)
$$
So that
$$
\Xi^\o_{\La}(\b,\l)= \int_{\O_\L}d\m_\l(\vec x)e^{ -\b \big[v(\vec x)- E^{v}_\L(\vec x,\o)\big]}\Eq(short)
$$
Again, we are supposing that  $\R^d$ is partitioned in elementary cubes $\D_\d$ of size $\d>0$ with $\d$ chosen in such a way that, for fixed cube $\L$ of size $2L$ centered at the origin,  $|\L_\d|/\d^d$ is integer so  that
$\L_\d$ denotes the set of elementary cubes forming $\L$. We
let  $\O_{\L_\d}^{1}$ to denote
the set of configurations
$\vec x\in \L$ such that in each cube $\D_\d\in \L_\d$
there is  one and only one particle. We   define the following crucial quantity.
$$
S^\o_\L(\d)=\sup_{\vec x\in \O_{\L_\d}^{1}}  E^{v^-}_\L(\vec x,\o)\Eq(sl)
$$
Note that, if $\o\in\O^*_g$ then by Lemma \ref{p2}  $S^\o_\L(\d)$ is well defined since it is bounded from above by $|\L_\d|\tilde \k
(1+g(L))$. Note also that if  $S^\o_\L(\d)=0$, then $E^{v^{_{-}}}_\L( x,\o)= 0$ for all $x\in \L$,
therefore $E^{v}_\L(\vec x,\o)=E^{v^+}_\L(\vec x,\o)$ for all $\vec x\in \L$ and hence
$$
\Xi^\o_{\La}(\b,\l)=\int_{\L}d\vec x e^{-\b v(\vec x)- \b
E^{v^+}_\L(\vec x,\o)}\le \int_{\L}d\vec x e^{-\b v(\vec x)}= \Xi^\0_{\La}(\b,\l)
$$
which implies trivially \equ(limsup).
Therefore we may suppose without loss  of generality that
$$
\liminf_{\L\to \infty}S^\o_\L(\d)>0\Eq(sig)
$$

\\Let us also define
$$
K^\o_\L=\sup_{x\in \L} E^{v^-}_\L(x,\o)
$$
Note that
$$
S^\o_\L(\d)>0 ~~\Longleftrightarrow~~ K^\o_\L>0
$$
and, due to definition of $S^\o_\L(\d)$, we have,
for any $\d>0$, that
$$
S^\o_\L(\d)\ge K^\o_\L\Eq(SK)
$$
Moreover,  via Lemma \ref{p2},  we can bound
$$
K^\o_\L \le \tilde  \k (1+g({L})) \Eq(Kbo)
$$

\\We will begin the proof of Theorem \ref{sergiosup} by proving below, as a consequence of the assumed superstability of the pair potential $v$, a key lemma (Lemma \ref{sergiolem} below).
Guessing that the statement  of this lemma may sound rather technical, we anticipate, before enunciating it, its interpretation  and its purpose. If
we have a configuration
of particles inside $\L$ that feels a strong negative energy from the outside
particles (measured by  the quantity $pS^\o_\L(\d)$ where $p$ is an integer), then this configuration must be constituted  by a
large number of particles and thus there are many pairs of
particles at short distance.
Lemma \ref{sergiolem} below shows the  contribution  to the energy of this large number of short-distance pairs of particles inside $\L$ is strongly   positive (i.e. of the order $p^2  S^\o_\L(\d)/K^\o_\L$). This positive energy, as will be shown later on,  is more than enough to compensate the effect from the outside particles, so that this kind of configurations will have low probability density and thus will be under control.

\begin{lema}\label{sergiolem} Let $\d\in (0, {a/\sqrt{d}})$. Given a potential $v$ as
in the theorem \ref{sergiosup}, let
$p\in \mathbb{N}$, $\vec x$ a configuration in a box $\L$ and $\omega \in \Omega$ such
that $K^\o_\L > 0$ and $E^{v^{_{-}}}_\L(\vec x,\o)> p S^\o_\L(\d)$. Then
$$
v_2(\vec x) \ge {c\over 4}p(p-1){S^\o_\L(\d)\over K^\o_\L}
$$
where $a$ and $c$ are  the constants appearing in \equ(deca).
\end{lema}

\\{\bf Proof}.
Due to definition \equ(sl), if $E^{v^-}_\L(\vec x,\o)> p S^\o_\L(\d)$, then
there exists at least a cube $\D_\d\in \L_\d$ containing $p+1$ particles.
Indeed if $\vec x$ is a configuration with at
most $p$ particles in each cube  then $ E^{v^-}_\L(\vec x,\o)\le p
S^\o_\L$ in contradiction with the hypothesis. Since $E^{v^-}_\L(\vec x,\o)>p S^\o_\L>(p-1)S^\o_\L$
then for the same reason we can find a
cube $\D_\d^1$ containing  at least $p$ particles of the configuration
$\vec x$ and, since $\d<a/\sqrt{d}$, all these particles in $\D_\d^1$ are at mutual distance less the $a$. Choose one particle inside $\D_\d^1$, call $x_1$ its position
and call $\vec x_1=\vec x\setminus \{x_1\}$. We have that
$$
v_2(\vec x) = \sum_{x\in \vec x_1}v_2(x-x_1)+ v_2(\vec x_1)\ge c(p-1)+v_2(\vec
x_1)
$$
Remove now $x_1$ from $\vec x$ so that we are left with the new
configuration $\vec x_1$. This new configuration is such that
$$
E^{v^-}_\L(\vec x_1,\o)= E^{v^-}_\L(\vec x,\o)-E^{v^-}_\L( x_1,\o)> pS^\o_\L(\d)-K^\o_\L
\ge   pS^\o_\L(\d)- S^\o_\L(\d)= (p-1)S^\o_\L(\d)
$$
So we could extract at least a point from the configuration $\vec x$ and yet, for the new configuration $\vec x_1$, the condition $E^{v^-}_\L(\vec x_1,\o)> (p-1)S^\o_\L(\d)$ still holds.  We can therefore repeat the process and extract $m\ge 1$ points from the
configuration $\vec x$ in such way that
$$
pS^\o_\L(\d)-mK^\o_\L>(p-1)S_\L^\o(\d)
$$
i.e. $m$ must be such that
$$
1\le m<{S_\L^\o(\d)\over K^\o_\L}
$$
Namely, we can extract
$$
m= \left\lfloor{S^\o_\L(\d)\over K^\o_\L}\right\rfloor
$$
points from  the configuration $\vec x$ in such  way that for the
remaining configuration $\vec x\,'=\vec x\setminus\{x_1,\dots x_m\}$
it holds
$$
E^{v^-}_\L(\vec x',\o)> (p-1)S^\o_\L(\d)
$$
and
$$
v_2(\vec x) \ge c\left\lfloor{S^\o_\L(\d)\over K^\o_\L}\right\rfloor(p-1)+v_2(\vec x\,')\ge
 c\left\lfloor{S^\o_\L(\d)\over K^\o_\L}\right\rfloor(p-1)+v_2(\vec x\,')
$$
Now the remaining configuration $\vec x\,'$  has the property  $ E^{v^-}_\L(\vec x\,',\o)> (p-1) S^\o_\L(\d)$. So, applying the same process
to bound  $v_2(\vec x\,')$ we get
$$
v_2(\vec x)\ge  c\left\lfloor{S^\o_\L(\d)\over K^\o_\L}\right\rfloor(p-1)+c\left\lfloor{S^\o_\L(\d)\over K^\o_\L}\right\rfloor(p-2)+ v_2(\vec x\,'')
$$
where now  $\vec x\,''$  is such that   $ E^{v^-}_\L(\vec x\,'',\o)> (p-2) S^\o_\L(\d)$.  Iterating we get
$$
v_2(\vec x) \ge c\left\lfloor{S^\o_\L(\d)\over K^\o_\L}\right\rfloor{p(p-1)\over 2}
$$
and since
$\left\lfloor{S^\o_\L(\d)\over K^\o_\L}\right\rfloor\ge {1\over 2}{S^\o_\L(\d)\over K^\o_\L}$ (because,
by \equ(SK),  $\left\lfloor{S^\o_\L(\d)\over K^\o_\L}\right\rfloor\ge 1$), the proof is concluded. $\Box$
\vskip.2cm
\\Another key ingredient of the proof is the following limit.

$$
\lim_{\L\uparrow\infty} {S^\o_\L(\d)K^\o_\L\over |\L|}=0  \Eq(limsl2)
$$
{\bf Proof of \equ(limsl2)}. We are supposing that $\L$ is a $d$ dimensional cube centered at the origin of size $2L$. We make a partition of $\Rd$ in elementary cubes $\D_\d$ of size $\d$ chosen in such a way that   $\L$ is formed by an integer number of elementary cubes and also in such a way that, for some constant $C_\d$, inequality \equ(kade) holds.

\\Recalling definitions \equ(evecpm) and  \equ(sl) we have
\begin{eqnarray*}
 S^\o_\L(\d) &=&  \sup_{\vec x\in \O_{\L_\d}^{1}}  E^{v^-}_\L(\vec x,\o) \\
   &\le & \sum_{\D_\d\subset \L}\sup_{x\in \D_\d}E^{v^-}_{{\L}}(x,\o)\\
   & \le & \sum_{\D_\d\subset \L}\sup_{x\in \D_\d}E^{v^-}_{d_x^\L}(x,\o)\\
 & \le  & \d^d \sum_{\D_\d\subset \L}\sup_{x\in \D_\d}\sum_{\D'_\d\in \L^c}\sup_{y\in \D'_\d}\eta(\|x-y\|)\r^\o_\d(y)
\end{eqnarray*}
Now, similarly as we did in the proof of Lemma \ref{p2} we may use  inequality \equ(kade) to bound, for some constant $C_\d$,
$$
\d^{d}\sup_{ y\in \D'_\d}\eta(\|x-y\|)\r^\o_\d(y)\le C_\d \int_{y\in \D'_\d} \eta(\| x- y\|)\r^\o_\d( y)dy
$$
Hence
$$
S^\o_\L(\d)~\le~C_\d\sum_{\D_\d\subset \L}\sup_{x\in \D_\d}\int_{y\in \L^c}\eta(\|x-y\|)\r^\o_\d(y)dy \le
C_\d\sum_{\D_\d\subset \L}\int_{y\in \L^c}\sup_{x\in \D_\d}\eta(\|x-y\|)\r^\o_\d(y)dy
$$
Now,
we  again  bound $\d^d\sup_{x\in \D_\d}\eta(\|x-y\|)$ by $C_\d\int_{\D_\d}\h(\|x-y\|)dx$  and we get
$$
S^\o_\L(\d)~\le~{C^2_\d\over \d^d}\sum_{\D_\d\subset \L}\int_{y\in \L^c}\Big(\int_{\D_\d}\h(\|x-y\|)dx\Big)\r^\o_\d(y)dy =
{C^2_\d\over \d^d}\sum_{\D_\d\subset \L}\int_{\D_\d}dx\int_{y\in \L^c}\h(\|x-y\|)\r^\o_\d(y)dy
$$
$$
= {C^2_\d\over \d^d}\int_\L dx\int_{y\in \L^c}\h(\|x-y\|)\r^\o_\d(y)dy
$$
I.e.,  setting $K_\d= { C^2_\d\over \d^d}$,  we get
\begin{eqnarray}\nonumber
S^\o_\L(\d)& \le & { K_\d}\int_{x\in\L}dx\int_{y\in \L^c}\eta(\|x-y\|)\r^\o_\d(y)dy\\
\nonumber & \le & {\r  K_\d} \int_{x\in\L}dx\int_{y\in \L^c}\eta(\|x-y\|)(1+g(\|y\|))dy\\
\nonumber & \le & {\r  K_\d} \int_{x\in\L}dx\int_{y\in \L^c}\eta(\|x-y\|)\Big[1+g(\|x-y\|)+ g(\|x\|)\Big]dy
\end{eqnarray}
Now,
$$
\int_{y\in \L^c}\eta(\|x-y\|)\Big[1+g(\|x-y\|)+ g(\|x\|)\Big]dy\le  \int_{\|y\|\ge d_x^\L}\eta(\|y\|)\Big[1+g(\|y\|)+ g(\|x\|)\Big]dy
$$
where recall that $d^\L_x$ is the distance of $x\in \L$ from the boundary $\partial \L$ of $\L$. Moreover,
since $\sup_{x\in \L}g(\|x\|)=g(\sqrt{d}L)\le g(dL)\le dg(L)$ we can bound
\begin{eqnarray*}
\int_{y\in \L^c}\eta(\|x-y\|)[1+g(\|x-y\|)+ g(\|x\|)]dy & \le & \int_{\|y\|\ge d_x^\L}\eta(\|y\|)\Big[1+g(\|y\|)+dg(L)\Big]dy\\
& = &  W(d_x^\L)+(1+dg(L))V(d^\L_x)
\end{eqnarray*}
where in the last line we have used  definitions  \equ(er) and \equ(wr).
Therefore, setting
$$
F(d_x^\L)= W(d_x^\L)+(1+dg(L))V(d^\L_x)
$$
we have that
$$
S^\o_\L(\d)~\le ~{\r  K_\d} \int_{x\in\L}F (d_x^\L)dx
$$
Now, recalling that $\L$ is a $d$-dimensional hypercube of size $L$ centered at the origin and thus $0\le d_x^\L\le L/2$, we have that
$$
\int_{x\in\L}F (d_x^\L)dx = \int_0^{L\over 2}  F(r) 2d\Big[2\Big({L\over 2}-r\Big)\Big]^{d-1}dr \le 2dL^{d-1}  \int_0^{L}F(r)dr
$$
and thus  we have, for $L$ so large that $g(L)\ge 1$,
$$
S^\o_\L(\d)  ~ \le ~ {2d^2\r L^{d-1} K_\d}\int_{0}^{L} \Big[W(r)+(1+g(L))V(r)\Big]dr\le   {4d^2\r L^{d-1} K_\d}\int_{0}^{L}  \Big[W(r)+g(L)V(r)\Big]dr
$$

\\Now, by Lemma \ref{p2} we have, for $\L$ sufficiently large (so that $g(L)>1$)
$$
K^\o_\L\le \tilde \k g(L)
$$
Therefore,   since $|\L|=(2L)^d\ge 2L^d$,  we have, setting  $\k_\d\le {2d^2\r L^{d-1} K_\d\tilde \k}$,
$$
 {S_\L^\o(\d)K_\L^\o\over |\L|}\le \k_\d\left[g(L) {\int_{0}^L  W(r)dr\over L}+ [g(L)]^2{\int_{0}^LV(r)dr\over L} \right]\Eq(rhs)
$$
and thus, given that $g$ satisfies \equ(diffi), \equ(limsl2) is proved. $\Box$

\vskip.2cm
\\We are now in the position to prove 
the Theorem \ref{sergiosup}. 
Let set
$$
E_\L= [S_\L^\o(\d)K^\o_\L]^{1\over 3}|\L|^{2\over 3}\Eq(ella)
$$
By \equ(limsl2) we have that
$$
\lim_{\L\to \infty } {{E_\L}\over |\L|}=\lim_{\L\to \infty } \left({S_\L^\o(\d)K_\L^\o\over |\L|}\right)^{1\over 3}=0 \Eq(flsl0)
$$
$$
\lim_{\L\to \infty } {{E_\L}\over S^\o_\L(\d)K^\o_\L}=\lim_{\L\to \infty } \left({|\L|\over S^\o_\L(\d)K_\L^\o}\right)^{2\over 3}=+\infty \Eq(flsl)
$$
We now can  write
$$
\Xi^\o_{\La}(\b,\l)=\int_{\O_\L}d\m_\l(\vec x)e^{-\b v(\vec x)- \b E^{v}_\L(\vec x,\o)}
~~~~~~~~~~~~~~~~~~~~~~~~~~~~~~~~~~~~~~~~~~~~~~~~~~~~~~~~~~~~~~~~~~~~~~~~~~~~~~~~~~~~~
$$
$$~~~~~~~\,\, =~
\int_{\O_\L:\, E^{v^-}_\L(\vec x,\o)\le E_\L}d\m_\l(\vec x)e^{-\b v(\vec x)- \b E^{v}_\L(\vec x,\o)}~+~
\int_{\O_\L:\,E^{v^-}_\L(\vec x,\o)>E_\L}d\m_\l(\vec x)e^{-\b v(\vec x)- \b E^{v}_\L(\vec x,\o)}
$$
$$~~~~\le~
e^{\b E_\L}\int_{\O_\L:\, E^{v^-}_\L(\vec x,\o)\le E_\L}d\m_\l(\vec x)e^{-\b v(\vec x)}~+~
\int_{\O_\L:\,E^{v^-}_\L(\vec x,\o)>E_\L}d\m_\l(\vec x)e^{-\b v(\vec x)+ \b E^{v^-}_\L(\vec x,\o)}
$$
$$\le~
e^{\b E_\L}\int_{\O_\L}d\m_\l(\vec x)e^{-\b v(\vec x)}~+~
\int_{\O_\L:\, E^{v^-}_\L(\vec x,\o)>E_\L}d\m_\l(\vec x)e^{-\b [v(\vec x)- E^{v^-}_\L(\vec x,\o)]}
~~~~~~~~~~~
$$
Namely, we get
$$
\Xi^\o_{\La}(\b,\l) ~\le~ e^{ \b
{E_\L}}\Xi^\0_{\La}(\b,\l)~+~
\int_{\O_\L~:\, E^{v^-}_\L(\vec x,\o)>E_\L}d\m_\l(\vec x)e^{-\b [v(\vec x)- E^{v^-}_\L(\vec x,\o)]}\Eq(terms)
$$
Let us consider the second term in the r.h.s. of inequality \equ(terms). By
hypothesis $v=v_1 +v_2$ with $v_1$ stable with stability constant equal to $B$.  Therefore  we can bound
$$
v(\vec x)-E^{v^-}_\L(\vec x,\o)~\ge~ -B|\vec x|~+~ v_2(\vec x)\,-~E^{v^-}_\L(\vec x,\o)
$$
We can now use  Lemma \ref{sergiolem} to bound from below $v_2(\vec x)-E^{v^-}_\L(\vec x,\o)$.
Let  $p$ be defined as the following integer.
$$
p=\left\lfloor {E^{v^-}_\L(\vec x,\o)\over 2S^\o_\L(\d)}\right\rfloor +2
$$
By the fact that we are considering here second term in the r.h.s. of inequality \equ(terms) where  $E^{v^-}_\L(\vec x,\o)>E_\L$ and  since \equ(flsl)  implies that
${E_\L/S^\o_\L(\d)}$ goes to infinity when $\L\uparrow \infty$, we have that  ${E^{v^-}_\L(\vec x,\o)/S^\o_\L(\d)}$ is surely larger than 4 for $\L$ large enough.
Then, using that ${x}\ge \lfloor {x\over 2}\rfloor+2$ for all $x\ge 4$, we have
$$
E^{v^-}_\L(\vec x,\o)= {E^{v^-}_\L(\vec x,\o)\over S^\o_\L(\d)}S^\o_\L(\d) >
\left(\left\lfloor {E^{v^-}_\L(\vec x,\o)\over 2S^\o_\L(\d)}\right\rfloor+2\right) S_\L(\d)^\o=p S^\o_\L(\d)
$$
Hence we can use Lemma \ref{sergiolem} to bound
$$
v_2(\vec x)-E^{v^-}_\L(\vec x,\o)~\ge~ {c\over
4}p(p-1){S^\o_\L(\d)\over K^\o_\L}-E^{v^-}_\L(\vec x,\o)
~~~~~~~~~~~~~~~~~~~~~~~~~~~~~~~~~~~~~~~~~~~~~~
$$
$$
~~~~~~~~~~~~~~~~~~~~~~= ~{c\over 4}\left( \left\lfloor {E^{v^-}_\L(\vec x,\o)\over 2S^\o_\L(\d)}\right\rfloor+2\right)
 \left(\left\lfloor {E^{v^-}_\L(\vec x,\o)\over 2S^\o_\L(\d)}\right\rfloor +1\right){S^\o_\L(\d)\over K^\o_\L}-E^{v^-}_\L(\vec x,\o)
$$
$$
~~~~~~~~~\ge~ {c\over 4}\left( {E^{v^-}_\L(\vec x,\o)\over 2S^\o_\L(\d)}\right)^2
 {S^\o_\L(\d)\over K^\o_\L}-E^{v^-}_\L(\vec x,\o)
 ~~~~~~~~~~~~~~~~~~~~~~~
$$
$$
= ~ E^{v^-}_\L(\vec x,\o)\left[{c\over 16}
{E^{v^-}_\L(\vec x,\o)\over S^\o_\L(\d)K^\o_\L} -1\right]
~~~~~~~~~~~~~~~~~~~~~~
$$
$$
\ge~ E_\L\left[{c\over 16}
{E_\L\over S^\o_\L(\d)K^\o_\L} -1\right]
~~~~~~~~~~~~~~~~~~~~~~~~~~~~~~~~
$$
where in the last line we have once again considered that  we are bounding the second term
in r.h.s. of \equ(terms) in which the integral is over  configurations
$\vec x$  such that $E^{v^-}_\L(\vec x,\o)\ge E_\L$.

\\In conclusion we have obtained that
$$
v_2(\vec x)-E^{v^-}_\L(\vec x,\o) \ge E_\L\left[{c\over16}  {E_\L\over
[S^\o_\L(\d)K_\L^\o] } -1\right] \doteq G_\L
$$
Let us analyze the behaviour of the ratio ${G_\L/|\L|}$ as $\L\to \infty$.  Recalling \equ(ella) and  \equ(Kbo), we get
$$
{G_\L\over |\L|}\ge {E_\L\over |\L|}\left[{c\over 16}  {E_\L\over
S^\o_\L(\d) K_\L^\o}-1\right] =
{c\over 16}\left( {|\L|\over S^\o_\L(\d)K_\L^\o }\right)^{1\over 3} - \left({S_\L^\o(\d)K_\L^\o\over |\L|}\right)^{1\over 3}
$$
and thus in force of \equ(flsl0) and \equ(flsl)  we have that
$$
\lim_{\L\to \infty} {G_\L\over |\L|}=+\infty
$$
\\Therefore
$$
\Xi^\o_{\La}(\b,\l)~\le ~e^{ \b
{E_\L}}\Xi^\0_{\La}(\b,\l)~+~
\int_{\O_\L\atop E^{v^-}_\L(\vec x,\o)>E_\L}d\m_\l(\vec x)e^{-\b [v(\vec x)- E^{v^-}_\L(\vec x,\o)]}
~~~~~~~~~~~~~~~~~~~~~~
$$
$$
\le ~e^{
\b {E_\L}}\Xi^\0_{\La}(\b,\l) ~+ ~e^{-\b G_\L}\int\limits_{\L}d\m_\l(\vec x) e^{+\b B|\vec x|}
~~~~~~~~~~~~~~~~~~~~~~~~~~~
$$
$$
\le ~e^{
\b {E_\L}}\Xi^\0_{\La}(\b,\l)~+~e^{-\b G_\L} e^{+\l|\L|e^{\b B}}
~~~~~~~~~~~~~~~~~~~~~~~~~~~~~~~~~~~~~~
$$
$$
\le~~ e^{-|\L|({\b G_\L\over |\L|}- \l e^{\b
B})} ~+~ e^{ \b {E_\L}}\Xi^\0_{\La}(\b,\l)
~~~~~~~~~~~~~~~~~~~~~~~~~~~~~~~~~~~~\,
$$
and thus
$$
{1\over |\L|}\log \Xi^\o_{\La}(\b,\l)~\le~ {1\over
|\L|}\log\left[e^{-|\L|({\b G_\L\over |\L|}- \l e^{\b B})} ~+~ e^{ \b
{E_\L}}\Xi^\0_{\La}(\b,\l)\right]
$$
In conclusion, we get
\begin{eqnarray*}
 \limsup_{\L\uparrow\infty}{1\over |\L|}\log \Xi^\o_{\La}(\b,\l) &\le & \lim_{\L\uparrow\infty}{1\over
|\L|}\log\left[e^{-|\L|({\b G_\L\over |\L|}- \l e^{\b B})} ~+ ~e^{ \b
{E_\L}}\Xi^\0_{\La}(\b,\l)\right] \\
   &= & \lim_{\L\uparrow\infty}{1\over
|\L|}\log\left[ + e^{ \b
{E_\L}}\Xi^\0_{\La}(\b,\l)\right] ~+ ~\lim_{\L\uparrow\infty}{1\over|\L|} \log\left(1+ {e^{-|\L|({\b G_\L\over |\L|}- \l e^{\b B})}\over e^{ \b
{E_\L}}\Xi^\0_{\La}(\b,\l)}\right) \\
 &\le& \lim_{\L\uparrow\infty}{1\over
|\L|}\log\left[ e^{ \b
{E_\L}}\Xi^\0_{\La}(\b,\l)\right]~ +~ \lim_{\L\uparrow\infty}{1\over|\L|} \log\left(1+ {e^{-|\L|({\b G_\L\over |\L|}- \l e^{\b B})}}\right) \\
 &= & \lim_{\L\uparrow\infty}{\b E_\L\over
|\L|}~+~ \lim_{\L\uparrow\infty}{1\over
|\L|}\log\Xi^\0_{\La}(\b,\l) \\
&= & \lim_{\L\uparrow\infty}{1\over
|\L|}\log\Xi^\0_{\La}(\b,\l)
\end{eqnarray*}
and thus inequality \equ(limsup) is proved. This concludes the proof of Theorem \ref{sergiosup}.

\section{Proof of Theorem \ref{sergioinf}}
We start by proving the following preliminary lemma.
\begin{lema}\label{p2+}
Let $g$ be admissible and let $\o\in \O^*_g$.
Then there exists  a  finite constant $\bar \k$  such that, for any $x\in\L$ such that $d^\L_x\ge b$
$$
E^{v^+}_\L(x,\o)\le\bar \k\left[W(d^\L_x)+ (1+g(L))V(d^\L_x)\right]
$$
\end{lema}
{\bf Proof}. If $\o\in \O^*_g$, then there exists $\r\in [0, \infty)$ such that  $\r_\d^\o(y)\le \r g(\|y\|)$ for all $y\in \Rd$. Moreover,
given  $x\in \L$ such that  $d^\L_x\ge b$, we have that $v^+(x-y)\le \h(\|x-y\|)$ for any $y\in \L^c$ . Therefore   thus we can bound
$$
 E^{v^+}_{\L}(x,\o)~
 \le~ \sum_{\D_\d\subset \L^c} \sup_{y\in \D_\d}
 v^+(x-y)|\o\cap\D_\d| ~
 \le~
\d^d\sum_{\D_\d\subset \L^c}\sup_{y\in\D_\d}\eta(\|x-y\|)\r_\d^\o(y)
$$
As we did previously (see \equ(kade)),  we can find a  constant $C_\d$ such that
$$
\d^d\sup_{y\in\D_\d}\eta(\|x-y\|)\r_\d^\o(y)\le C_\d\int_{\D_d}\eta(\|x-y\|)\r_\d^\o(y)dy
$$
Therefore
\begin{eqnarray*}
E^{v^+}_{\L}(x,\o) & \le & C_\d\sum_{\D_\d\subset \L^c} \int_{\D_d}\eta(\|x-y\|)\r_\d^\o(y)dy \\
   & = & C_\d\int_{ \L^c}\eta(\|x-y\|)\r_\d^\o(y)dy \\
   &\le & C_\d\r\int_{ \L^c}\eta(\|x-y\|)(1+g(\|y\|))dy \\
   & \le & C_\d\r\int_{\L^c}\eta(\|x-y\|)(1+g(\|x-y\|+\|x\|))dy
\end{eqnarray*}
Now, using again \equ(subli), we get
$$
g(\|x-y\|+\|x\|)\le [g(\|x-y\|)+ g(\|x\|)]
$$
and therefore
\begin{eqnarray*}
E^{v^+}_{{\L}}(x,\o) &\le & C_\d\r\int_{\L^c}\Big[1+g(\|x-y\|)+ g(\|x\|)\Big]\eta(\|x-y\|) dy\\
 &\le & C_\d \r \left[\int_{\L^c}{g(\|x-y\|)\h(\|x-y\|)}dy+ (1+g(\|x\|))
\int_{\L^c}{\h(|x-y\|)}dy\right] \\
  &\le & C_\d \r  \left[\int_{\|x-y\|\ge d^\L_x}{g(\|x-y\|)\h(\|x-y\|)}dy~+~ (1+g(\|x\|))
\int_{\|x-y\|\ge d^\L_x}{\h(|x-y\|)}dy\right]\\
 & \le & C_\d \r d\left[W(d^\L_x)+ (1+g(L))V(d^\L_x)\right]
\end{eqnarray*}
where in the last line   we have again used definitions \equ(er) and \equ(wr) and  the fact that $g(\|x\|)\le dg(L)$  for any  $x\in \L$. $\Box$

\vskip.2cm
\\Using  Lemma \ref{p2+}  we can now    conclude the proof of  Theorem \ref{sergioinf}.
By  hypothesis there exists an increasing continuous function
$h(L)$ such that   $\lim_{L\to \infty} h(L)=\infty$,  $\lim_{L\to \infty} h(L)/L=0$ and
$$
\lim_{L\to \infty} g(L) V(h(L))=0\Eq(glvl)
$$
We take $L$ sufficiently   large in such a way that $b<h(L)<L$, and define $\L_h=\{x\in \L: d_x^\L>h(L)\}$ so that
 $\L_h$ is a cube centered at the origin with size $2(L-h(L))$ fully contained  in $\L$.
Therefore we have that

$$
\Xi_\L^\o(\b\l)\ge\int_{\O_{\L_h}}d\m_\l(\vec x)e^{-\b v(\vec x)- \b E^{v}_\L(\vec x,\o)}\ge
\int_{\O_{\L_h}}d\m_\l(\vec x)e^{-\b v(\vec x)- \b E^{v^+}_\L(\vec x,\o)}
$$
Now by definition, for all $x\in \L_h$ we have that $d^\L_x\ge h(L)>b$   and thus we can apply Lemma  \ref{p2+}
to bound, for any $x\in \L_h$
$$
 E^{v^+}_\L(x,\o)\le \tilde \k \Big[W(h(L))+ [1+g(L)]V(h(L))\Big]
$$
Moreover,  since,    by \equ(vrinfty), \equ(wrinfty) and \equ(glvl),
$\lim_{\L\uparrow\infty} \big[ W(h(L))+ [1+g(L)]V(h(L))\big]=0$,
for $\L$ large enough and for any fixed $\e>0$, we can bound  $E^{v^+}_\L(x,\o)\le\e$ so that
$$
\Xi_\L^\o(\b\l)\ge
\int_{\O_{\L_h}}d\m_\l(\vec x)e^{-\b v(\vec x)- {\b \e} |\vec x|}= \Xi_{\L_h}^\0(\b,e^{-\b \e}\l)
$$
Therefore,  considering that $\lim_{\L\uparrow\infty}\L_h=+ \infty$ and that $\lim_{\L\uparrow\infty} {|\L_h|\over  |\L|}=1$, we get
\begin{eqnarray*}
\liminf_{\L\uparrow\infty} {1\over |\L|}\log \Xi_\L^\o(\b,\l) &\ge & \lim_{\L\uparrow\infty} {1\over |\L|}
\log \Xi_{\L_h}^\0(\b,e^{-\b \e}\l) \\
   &= & \lim_{\L\uparrow\infty} {|\L_h|\over |\L|} {1\over |\L_h|}  \log\Xi_{\L_h}^\0(\b,e^{-{\b \e}}\l) \\
  &=& \lim_{\L\uparrow\infty} {|\L_h|\over |\L|} \lim_{\L\uparrow\infty} {1\over |\L_h|}
\log \Xi_{\L_h}^\0(\b,e^{-\b \e}\l) \\
   &=& \lim_{\L_h\uparrow \infty}{1\over |\L_h|} \Xi^\0_{\L_h}(\b,e^{-\b \e}\l) \\
   &=& \b p^\0(\b,e^{-\b \e}\l)
\end{eqnarray*}
Now, since the free-boundary condition infinite volume pressure $p^\0(\b,\l)$ is continuous as a function of $\b$ and $\l$, by the arbitrariness of $\e$  we can conclude that,
$$
\liminf_{\L\uparrow\infty} {1\over |\L|}\Xi_\L^\o(\b,\l)~\ge~ \b p^\0(\b,\l)~= ~\lim_{\L\uparrow \infty}{1\over |\L|} \Xi^\0_{\La}(\b,\l)\Eq(Okkk)
$$
This ends the proof of Theorem \ref{sergioinf}.


\section{Conclusions}
In this note we considered  a $d$-dimensional system of classical particles confined in a cubic box  $\L$ interacting  via
a superstable pair potential in the Grand Canonical ensemble at fixed inverse temperature $\b>0$ and fixed fugacity $\l>0$. We proved that the thermodynamic limit of the finite volume pressure of such system
does not depend on boundary conditions generated
by particles at fixed positions outside the volume $\L$ as long as these external particles  are distributed according to a bounded density
$\r_{\rm ext}$ (even larger as we please  than the density  $\r_0(\b,z)$  of the system calculated using  free boundary conditions). We also prove the independency of  the thermodynamic limit of the  pressure of the system in presence of boundary conditions whose
density may increase with the distance from the origin to a rate which depends on how fast the pair potential decays.

\\A related  open question (and possibly the subject of a project  to come) is whether it  is possible to perform an absolutely convergent Mayer expansion of the pressure  of the systems considered in this note (i.e. interacting via a non-necessarily repulsive pair potential) for
fugacities within a convergence radius  uniform in the boundary conditions when
these are in the class described above.

\section*{Acknowledgments}
A.P.  has been partially supported by the Brazilian  agencies
Conselho Nacional de Desenvolvimento Cient\'{\i}fico e Tecnol\'ogico
(CNPq  - Bolsa de Produtividade em pesquisa, grant n. 306208/2014-8)
and Coordena\c{c}\~ao de Aperfei\c{c}oamento de Pessoal de N\'\i vel Superior
(CAPES - Bolsa PRINT, grant n. 88887.474425/2020-00). S.Y. has been partially supported by the Argentine agency
CONICET (Consejo Nacional de Investigaciones Cient\'\i ficas y
T\'ecnicas).

\end{document}